\documentclass{revtex4}

\begin{document}

\markboth{} {\textit{Effect of a Magnetic Field on the Electroweak
Symmetry}}

%
%

\title{EFFECT OF A MAGNETIC FIELD ON THE ELECTROWEAK SYMMETRY}

\author{E. RODR\'{I}GUEZ QUERTS}

\address{ICIMAF, Calle E No. 309, La Habana,10400, Cuba\\
elizabeth@icmf.inf.cu}

\author{H. P\'{E}REZ ROJAS}

\address{ICIMAF, Calle E No. 309, La Habana,10400, Cuba\\
hugo@icmf.inf.cu}

\author{A. P\'{E}REZ MART\'{I}NEZ}

\address{ICIMAF, Calle E No. 309, La Habana,10400, Cuba\\
aurora@icmf.inf.cu}


\begin{abstract}
We discuss the effect of a strong magnetic field in the behavior
of the symmetry of an electrically neutral electroweak plasma. We
analyze the case of a strong field and low temperatures as
compared with the W rest energy. If the magnetic field is large
enough, it is self-consistently maintained. Charged vector bosons
play the most important role, leading only to a decrease of the
symmetry breaking parameter, the symmetry restoration not being
possible.
\end{abstract}

\maketitle

\section{Introduction}

The Standard Model of the electroweak interaction at finite
temperatures predicts the existence of two phases : the symmetric
and the broken one at temperatures, respectively, above and below
some critical value $T_c$ \cite{sm}. The possibility of symmetry
restoration under a large magnetic field for this model was
firstly considered by Linde \cite{linde}, who observed that at
zero temperature, an increase of the field increases the symmetry
breaking parameter. He notices that only the vector particle
contribution may lead to the symmetry restoration at sufficiently
large field.

The characteristic field, which can substantially modify the
symmetry breaking parameter was estimated to be of the order of
$B_{ch}\sim 10^{27}G$. Later Ambjorn and Olesen \cite{ao1}
realized that the usual electroweak vacuum become unstable for
some critical field value $B_c=\frac{m_w^2}{e}\sim 10^{24}G$, due
to the presence of charged vector bosons W. They also obtained a
static magnetic solution of the classical electroweak equations,
corresponding to a vacuum condensate of W and Z bosons, for $
B>B_c $ \cite{ao2}.

The problem was considered in \cite{h3} from the quantum statistical point
of view. It was concluded that for $B\rightarrow B_c,$ the population of the
W-ground state increases, leading to a self-magnetization of the system;
actually this prevents $B$ to reach the critical value $B_c.$

We explore  the possibility that $m_w^2$ would decrease, via
$\xi$, with increasing $B$. We keep in mind the analogy with
superconductivity or either, between the symmetry breaking
parameter $\xi $ and a Bose-Einstein condensate. The condensate is
destroyed by a sufficiently large magnetic field (the critical
Schafroth field \cite{sch}). We will conclude in the present case
that there is actually a decrease of $\xi=\xi(B) $ for increasing
$B$, but we find that the decrease is a small fraction of $\xi
(0)$.

We study our problem in the frame of quantum statistics, taking in
mind possible consequences for astrophysics and cosmology.  We
consider the lepton sector of an electrically neutral electroweak
plasma. We evaluate the variation of the symmetry breakdown
parameter in the external field and examine the possibility of
symmetry restoration. The present paper is a summarized and
up-to-date version of the previous one \cite{sym}.

\section{The electroweak plasma}

Thermodynamical properties of an electroweak plasma in a constant
magnetic field can be studied if we know the effective potential
associated to the system. We obtain this potential for the
leptonic sector of the plasma, in the one loop approximation,
starting from the Weinberg -Salam model \cite{h plasma}, and it
has the form
\begin{equation}
V=V_e+V_w+V_h+V_z+V_A+V_\nu+V_t,  \label{pot}
\end{equation}
where the first two terms are related with charged particles
(electrons and W bosons). The terms due to the neutral vector
boson Z , the electromagnetic field and Higgs scalar are $V_z,V_A$
and $V_h$, respectively. The neutrinos contribution is $V_\nu$.
Finally, $V_t=\frac{\lambda _2}4\left( \frac{\xi ^2}2-a^2\right)
^2$
is the tree effective potential term, which depends on the symmetry breaking parameter $\xi $ ( $%
\lambda_2$ is the scalar coupling constant and $a$ is the "negative mass"
parameter ). The particle masses are related to the mentioned parameter by
the usual expressions $m_w=\frac g2\xi ,m_z=\frac 12\sqrt{g^2+g^{\prime 2}}\xi ,m_{\sigma}=\sqrt{\frac{
\lambda _2}2}\xi ,m_e=\lambda _1\xi$, where
 $g$, $g^{\prime },\lambda _1$ and $\lambda _2$ are, respectively, the electroweak,
Yukawa and Higgs scalar coupling constants. The value of the symmetry breaking
parameter is obtained from the extremum condition, determining
the temperature-dependent mass shell,
\begin{equation}
\partial V/\partial \xi =0.  \label{sym.c}
\end{equation}

By evaluating the effective potential on the mass shell, we get
the thermodynamical potential $V(\xi _{\min })=\Omega $. One can
write then two other equilibrium equations. One of them is the
lepton number conservation $-\partial \Omega /\partial
\mu_2=N_l=N_e+N_\nu$ (where $N_i$ is the net density of particles
(particles minus antiparticles), per unit volume) and the other
$\partial \Omega /\partial \mu_1=0$, is the electric charge
conservation, which in our simplified model is reduced to
$N_e+N_w=0$. The magnetization $M=-\partial \Omega/\partial B$
contains the contributions of both electrons and W bosons
$M=M_e+M_w$.

\section{The strong magnetic field limit: symmetry analysis}

In the absence of field and at zero temperature, the effective
potential coincides with the tree term $V\mid _{T=0,B=0}=V_t$, and
therefore the equation (\ref{sym.c}) has only one stable solution
$\xi _o=\sqrt{2}a$. It is known that temperature modify the
symmetry breaking parameter. In fact, the Higgs model predicts
\cite{linde} that an increase of temperature decreases the
symmetry breaking parameter and at some critical temperature $T_c$
the symmetry is restored ($T_c\sim 10^{15}K$ \cite{linde}). We can
expect then that an intense external magnetic field also modifies
the symmetry breaking parameter.

 We will restrict ourselves to the case of a strong
magnetic field and/or law temperatures, when the condition $eB\gg T^2$ is satisfied.
 It
can be demonstrated that, in our case, only the charged boson
contribution may substantially modify the symmetry breaking
parameter (see \cite{sym}). For $eB\gg T^2$ , the average W boson
population in excited Landau states is negligible small. Moreover,
in that limit the Bose-Einstein distribution degenerates in a
Dirac $\delta $ function and the most of the W density is in the
Landau ground state $n=0$ and distributed in a very narrow
interval around $p_3=0$ \cite{h 2}. If we only consider the
contribution of the W boson sector \footnotetext[1]{For the
W-sector, the effective potential $V_w=V_w^{st}+V_w^o$, where
 the first term is the statistical part and the second one
is the Euler-Heisenberg vacuum term. We can consider $V_w\approx
V_w^{st}$ (see \cite{sym}). }$^{\footnotemark[1]}$, Eqs.
(\ref{pot}) and (\ref{sym.c}) takes the form \footnotetext[2]{We
write $\frac{\xi ^2}{\xi _o^2}=y, \frac{eB}{m_w^2}=\frac zy$,
where $y>0,z>0$ and $z<y$, because we are considering fields $B <
B_c$.} $^{\footnotemark[2]}$(with $C=\frac{gN_w}{\sqrt{2}\lambda
_2a^3}$)
\begin{equation}
V=\frac{\lambda _2a^4}4\left[ (y-1)^2+4C\sqrt{y-z}\right] ,  \label{wsym2}
\end{equation}
\begin{equation}
[(y-1)\sqrt{y-z}+C]\sqrt{y}=0. \label{wsymc2}
\end{equation}

It can be shown that for fields less than some value $B_m(N_w)$
(that is, for $z<z_m=1-3\sqrt[3]{2C^2}/2$), there is a non zero
symmetry breaking parameter $\xi $, which corresponds to the
minimum of the effective potential. When the field grows, $\xi $
decreases, and for fields equal or greater than $B_m$ ($z\geq
z_m$), the effective potential will not have a stable equilibrium
point.  That is, the broken solution symmetry is no longer true.
On the other hand, the  solution $\xi=0$ is unrealistic, leading
to purely imaginary physical quantities: i.e. if $B \neq 0$ and ,
we have $\varepsilon_0 (p_3 =0) =\sqrt{-eB}$.

Actually, for solving our problem, we must take into account that for sufficiently large magnetization ($%
M\gg H$) it can self-consistently maintain the field $B$. So, we
can put $ B=4\pi M$, where $M_w= eN_w/2\sqrt{m_w^2-eB}$,
 and consider this equation together
with Eq.(\ref{wsymc2}), and by calling $D =FC$, $F= 8\pi
e^2\lambda_2/g^4$ we have,
\begin{equation}
z\sqrt{y-z}-D=0. \label{selfmagn}
\end{equation}
 We may express the solutions $z$ and $y$ as functions of $x_i=\frac 2{\sqrt{3}}\cos \frac{\kappa _i\pi -\arctan \sqrt{\frac 4{27E^2}-1}%
}3$  (with $E=C+D$):
\begin{equation}
z_{1,2}=F\frac{1-x_{1,2}^2}{1+F}, \hspace{1cm}
y_{1,2}=\frac{F+x_{1,2}^2}{1+F}. \label{soly}
\end{equation}
where $i=1,2;$ $\kappa _i=5,1$ respectively, and $0\leq x_1\leq 1/\sqrt{3},1/%
\sqrt{3}\leq x_2\leq 1.$. Due to the fact that $E$ is proportional to $N_w$, we deduce that for each $%
N_w$ , there are two possible values of the symmetry breaking parameter $%
y_{1,2}$. It is easy to prove that $y_1$ corresponds to an
unstable equilibrium point of the potential, while $V$ has a
minimum for $y=y_2$. We conclude, thus, that as $N_w$ grows, the
field also grows and the symmetry breaking parameter decreases to
the minimum $\xi_{\min }=\kappa \xi_o,$ where $\kappa
=\sqrt{(1+3F)/3(1+F)}$ and $ F= 4\pi sin^2 \theta_w \eta^2$, where
$\eta= m_\sigma /m_w$ and $\theta_w$ is the Weinberg angle. Thus,
$\kappa$$(\leq 1)$ increases with increasing $\eta^2$. We take for
the Higgs mass the lower bound of $114.4$ Gev, according to recent
estimates \cite{Hagiwara} from Particle Data Group. Then $\eta
\geq 1.42$, and one gets $F\geq 5.65$ , $1> \kappa \geq 0.949$,
which means a 5 per cent reduction of the symmetry breaking
parameter.
 The corresponding density is $N_w \simeq 4\cdot 10^{47}$.
Fixing $\eta$, any further increase of the $N_w$, would not lead
to real solutions of $V$. This means that excited $W$ Landau
states start to be populated. But these states contribute
diamagnetically to the total magnetization, and therefore, it is
kept $4\pi M=B<B_c$. Higher order corrections do not change the
essence of the symmetry behavior in the present problem
\cite{sym}.

\section{Conclusions}

We conclude that for a neutral electroweak plasma in
a large constant magnetic field,  only the charged vector particle contribution may
substantially modify the symmetry breaking parameter. For high values of the field, it is
maintained self-consistently and the field never reaches its critical
value $B_c$. The symmetry breaking parameter is decreased some amount
under the action of the magnetic field, and in consequence, the masses of
electrons, W and Z bosons, and Higgs particles become slightly smaller than
in the zero field case at zero temperature.

\end{document}